\documentstyle{mn}

\newif\ifAMStwofonts
\AMStwofontstrue

\input{epsf}

\def\ee{$e^\pm$}
\def\g{$\gamma$}
\def\ginga{{\it Ginga}}
\def\exosat{{\it EXOSAT}}
\def\asca{{\it ASCA}}

\ifoldfss
  \ifCUPmtlplainloaded \else
    \NewTextAlphabet{textbfit} {cmbxti10} {}
    \NewTextAlphabet{textbfss} {cmssbx10} {}
    \NewMathAlphabet{mathbfit} {cmbxti10} {} 
    \NewMathAlphabet{mathbfss} {cmssbx10} {} 
  \fi
  \ifAMStwofonts
    \ifCUPmtlplainloaded \else
      \NewSymbolFont{upmath} {eurm10}
      \NewSymbolFont{AMSa} {msam10}
      \NewMathSymbol{\upi}     {0}{upmath}{19}
      \NewMathSymbol{\umu}     {0}{upmath}{16}
      \NewMathSymbol{\upartial}{0}{upmath}{40}
      \NewMathSymbol{\leqslant}{3}{AMSa}{36}
      \NewMathSymbol{\geqslant}{3}{AMSa}{3E}

       \let\le=\leqslant
      \let\geq=\geqslant 
    \fi
  \fi
\fi 

\ifnfssone
  \newmathalphabet{\mathit}
  \addtoversion{normal}{\mathit}{cmr}{m}{it}
  \addtoversion{bold}{\mathit}{cmr}{bx}{it}
  \newmathalphabet{\mathbfit} 
  \addtoversion{normal}{\mathbfit}{cmr}{bx}{it}
  \addtoversion{bold}{\mathbfit}{cmr}{bx}{it}
  \newmathalphabet{\mathbfss} 
  \addtoversion{normal}{\mathbfss}{cmss}{bx}{n}
  \addtoversion{bold}{\mathbfss}{cmss}{bx}{n}
  \ifAMStwofonts
    \ifCUPmtlplainloaded \else
      \UseAMStwoboldmath
      \makeatletter
      \new@mathgroup\upmath@group
      \define@mathgroup\mv@normal\upmath@group{eur}{m}{n}
      \define@mathgroup\mv@bold\upmath@group{eur}{b}{n}
      \edef\UPM{\hexnumber\upmath@group}
      \new@mathgroup\amsa@group
      \define@mathgroup\mv@normal\amsa@group{msa}{m}{n}
      \define@mathgroup\mv@bold\amsa@group{msa}{m}{n}
      \edef\AMSa{\hexnumber\amsa@group}
      \makeatother
      \mathchardef\upi="0\UPM19
      \mathchardef\umu="0\UPM16
      \mathchardef\upartial="0\UPM40
      \mathchardef\leqslant="3\AMSa36
      \mathchardef\geqslant="3\AMSa3E

       \let\le=\leqslant
      \let\geq=\geqslant 
    \fi
  \fi
\fi 

\ifnfsstwo
  \DeclareMathAlphabet{\mathbfit}{OT1}{cmr}{bx}{it}
  \SetMathAlphabet\mathbfit{bold}{OT1}{cmr}{bx}{it}
  \DeclareMathAlphabet{\mathbfss}{OT1}{cmss}{bx}{n}
  \SetMathAlphabet\mathbfss{bold}{OT1}{cmss}{bx}{n}
  \ifAMStwofonts
    \ifCUPmtlplainloaded \else
      \DeclareSymbolFont{UPM}{U}{eur}{m}{n}
      \SetSymbolFont{UPM}{bold}{U}{eur}{b}{n}
      \DeclareSymbolFont{AMSa}{U}{msa}{m}{n}
      \DeclareMathSymbol{\upi}{0}{UPM}{"19}
      \DeclareMathSymbol{\umu}{0}{UPM}{"16}
      \DeclareMathSymbol{\upartial}{0}{UPM}{"40}
      \DeclareMathSymbol{\leqslant}{3}{AMSa}{"36}
      \DeclareMathSymbol{\geqslant}{3}{AMSa}{"3E}

       \let\le=\leqslant
      \let\geq=\geqslant 
    \fi
  \fi
\fi 

\ifCUPmtlplainloaded \else
  \ifAMStwofonts \else 
    \def\upi{\upi}
    \def\umu{\mu}
    \def\upartial{\partial}
  \fi
\fi

\title[Correlation between Compton reflection and X-ray slope]
{Correlation between Compton reflection and X-ray slope in Seyferts
and X-ray binaries}

\author[A. A. Zdziarski, P. Lubi\'nski and D. A. Smith]
{\parbox[]{7in} {Andrzej A. Zdziarski$^1$, Piotr Lubi\'nski$^2$ and David A.
Smith$^{3,4,5}$}\\
 $^1$N. Copernicus Astronomical Center, Bartycka 18, 00-716 Warsaw, Poland \\
$^2$Heavy Ion Laboratory, Warsaw University, Pasteura 5a, 02-093 Warsaw, Poland
\\
$^3$Laboratory for High Energy Astrophysics, NASA/GSFC, Code 662, Greenbelt, MD
20771, USA \\
$^4$Department of Astronomy, University of Maryland, College Park, MD 20742,
USA \\
$^5$Dept. of Physics, University of Leicester, University Road, Leicester
LE1 7RH, UK \\
}

\date{Accepted 1998 December 2. Received 1998 October 9}

\pagerange{\pageref{firstpage}--\pageref{lastpage}}
\pubyear{1999}

\begin{document}

\maketitle

\label{firstpage}

\begin{abstract}
We find a very strong correlation between the intrinsic spectral slope in
X-rays and the amount of Compton reflection from a cold medium in Seyfert AGNs
and in hard state of X-ray binaries with either black holes or
weakly-magnetized neutron stars. Objects with soft intrinsic spectra show much
stronger reflection than ones with hard spectra. We find that at a given
spectral slope, black-hole binaries have similar or more reflection than
Seyferts whereas neutron-star binaries in our sample have reflection consistent
with that in Seyferts. The existence of the correlation implies a dominant role
of the reflecting medium as a source of seed soft photons for thermal
Comptonization in the primary X-ray source. 
\end{abstract}


\section{INTRODUCTION}
\label{s:intro}

Recently, a paradigm has appeared to emerge according to which black-hole
binaries in the hard state show weak both Compton reflection (i.e.\
backscattering of X-rays from a surrounding cold medium) and associated
fluorescent Fe K$\alpha$ lines (as found by Gierli\'nski et al.\ 1997; \.Zycki,
Done \& Smith 1997, 1998; Zdziarski et al.\ 1998; Ebisawa et al.\ 1996) whereas
Seyfert-1 AGNs would universally show stronger reflection (Nandra \& Pounds
1994) and strong and broad Fe K$\alpha$ lines (Nandra et al.\ 1997). If this
were correct, it would certainly be of importance for our understanding of the
physics of X-ray sources in accreting compact objects.

We critically examine this paradigm based on available data. We concentrate on
the continuum spectral properties, deferring an analysis of a more complex
issue of the Fe K$\alpha$ fluorescent emission to a future work. Here, we study
the strength of Compton reflection as a function of the intrinsic spectral
slope. A correlation between these quantitites has originally been found in
\ginga\/ observations of GX 339--4 (Ueda, Ebisawa \& Done 1994). In this work,
we consider Seyferts, radio galaxies, and X-ray binaries containing either
black holes or weakly-magnetized neutron stars.

\section{THE OBSERVED CORRELATION}
\label{s:cor}

Our Seyfert sample consists of \ginga\/ (Makino et al.\ 1987) spectra
(extracted from the Leicester database) of radio-quiet (hereafter RQ) Seyfert
1s and narrow emission-line galaxies. The latter are Seyferts intermediate
between type 1 and 2 showing moderate X-ray absorption (e.g.\ Smith \& Done
1996). This sample is basically the same as that of the classical study of
\ginga\/ spectra of Seyferts of Nandra \& Pounds (1994). However, here we
exclude 3 radio-loud AGNs, as there are hints that their nature is different
from that of RQ ones (Wo\'zniak et al.\ 1998). On the other hand, we include
some late \ginga\/ observations of Seyfert 1s not listed in Nandra \& Pounds
(1994), which gives us 61 observations of 24 RQ Seyferts. We further include 2
\ginga\/ observations of 4U 0241+61, a low-redshift AGN ($z=0.044$), which we
find to be of RQ type after correcting its $B$ magnitude for Galactic
extinction. Finally, we include an observation of NGC 4151 contemporaneously by
both \ginga\/ and the OSSE detector on board of {\it CGRO}, which latter data
allow a determination of the strength of reflection in this strongly-absorbed
bright Seyfert (Zdziarski, Johnson \& Magdziarz 1996).

We compare the data for Seyferts with those for 4 X-ray binaries in the hard
(also called low) state. Two of them, Cyg X-1 and GX 339--4, are black-hole
candidates, and two, GS 1826--238 and 4U 1608--522, are X-ray bursters (thus
containing weakly-magnetized neutron stars). For Cyg X-1, we use 5, 3 and 4
observations from 1987, 1990 and 1991, respectively (Ebisawa et al.\ 1996 and
references therein; Gierli\'nski et al.\ 1997). For GX 339--4, we use 5
\ginga\/ observations out of 6 ones of Ueda et al.\ (1994) (excluding
one in an off state). A spectrum of GS 1826--238 (Strickman et al.\
1996) and 2 spectra of 4U 1608--522 from 1990 and 1991 (Yoshida et al.\ 1993)
have been extracted from the \ginga\/ database.

We use \ginga\/ spectra from both the top ($\sim 1.7$--20 keV) and mid ($\sim
10$--20 keV) layers of the LAC (Turner et al.\ 1989), to which data we add a
systematic error of 0.5 per cent per channel. Inclusion of the mid-layer data
significantly increases the effective area above 10 keV, allowing to measure
the hard X-ray spectra much more accurately than with the top layer alone (see
Wo\'zniak et al.\ 1998; Magdziarz et al.\ 1998). We further select only the
data with no systematic differences between the top and mid-layer spectra above
10 keV, which reduces the number of usable Seyfert observations by 10.

We fit the data (using {\sc xspec}, Arnaud 1996) with a continuum model
consisting of an e-folded power law (unless stated otherwise) and a component
due to its Compton reflection (Magdziarz \& Zdziarski 1995) at a normalization,
$R$. In the case of an isotropic primary source and no obscuration of either
the source or the reflector, $R=\Omega/2\upi$, where $\Omega$ is the solid
angle subtended by the reflector.  We initially assume the reflector is neutral
with the abundances of Anders \& Ebihara (1982), but allow for its ionization
and/or a free Fe abundance when it is statistically required. We fix the
e-folding energy at 400 keV (Zdziarski et al.\ 1995) and the reflector viewing
angle of Seyferts at $i=30\degr$ (Nandra et al.\ 1997). (For $i=45\degr$, the
fits below would typically give $R$ higher by $\sim 10$ per cent.) We model the
Fe K$\alpha$ line as a Gaussian with the line flux as a free parameter,
independent of reflection. This allows for resonant absorption and/or
additional line components due to either matter in the line of sight (Makishima
1986), scattering of a part of the primary continuum by an ionized medium
(Krolik \& Kallman 1987) and/or emission of a Thomson-thin torus surrounding an
AGN nculeus (e.g.\ Wo\'zniak et al.\ 1998).

Then we carefully treat the low-energy part of the spectra. For each object, we
initially model absorption of the above continuum by a neutral medium at the
redshift of the source (in addition to a fixed local absorber with the Galactic
column density). We then test for the effect of either replacing the neutral
absorber by an ionized one, ignoring some low-energy channels, or using a
broken power-law incident continuum. This is to assure that the hard X-ray
power-law photon index, $\Gamma$, is properly measured and not affected by
either the presence of a soft excess or by complex absorption. In the case of
NGC 4151, we assume $i=65\degr$ and fit the \ginga/OSSE data with a dual
absorber model (Zdziarski et al.\ 1996) and a thermal-Comptonization continuum
model (Misra et al., in preparation).

The above procedure gives us $R$ and $\Gamma$ with their error contours. For
Seyferts, some detections are rather weak and yield very large error contours.
Since our main goal is to study the dependence $R(\Gamma)$, we have eliminated
from the sample observations for which the 1-$\sigma$ ($\Delta\chi^2 = 2.30$)
error contour has $\Delta R >2$. Inclusion of such data would only weakly
affect our fitted dependences (see Section 3) but would confuse our graphical
representation of the results. This criterion eliminates, in particular, all
the observations of MCG --6-30-15 and Mkn 335. To improve statistics in those
cases, we have coadded multiple observations of each of those objects. In the
case of Mkn 335, the resulting $\Delta R$ was still $\gg 2$, and thus we kept
it out of the sample. For MCG --6-30-15, we coadded 4 (out of 5) observations
that were performed in the standard mode with all the LAC detectors operating,
which then resulted in data with an acceptable $\Delta R <2$.

\begin{figure*}
\label{fig:cor} \begin{center} \leavevmode \epsfxsize=17.6cm
\epsfbox{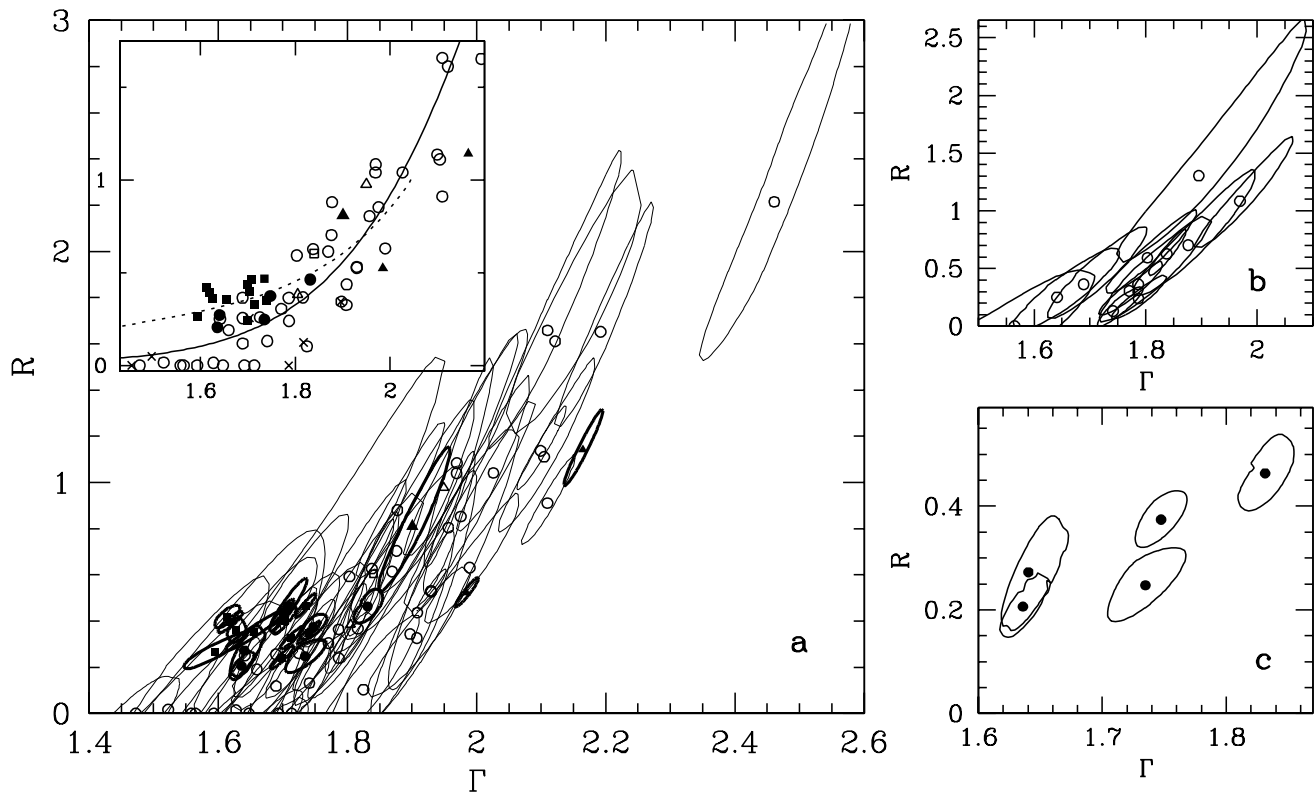}\end{center} \caption{The $R(\Gamma)$ correlation in
Seyferts and X-ray binaries in the hard state. (a) The data and models (curves
in the inset), see Sections 2 and 4, respectively. Examples of the correlation
for individual objects: (b) NGC 5548 and (c) GX 339--4. } \end{figure*}

The procedures outlined above give us 47 data sets for 23 RQ AGNs: 4U 0241+61,
AKN 120, Fairall 9, IC 4329A, MCG --2-58-22, MCG --5-23-16,  MCG --6-30-15, Mkn
509, Mkn 841, NGC 2110, NGC 2992, NGC 3227, NGC 3783, NGC 4051, NGC 4151, NGC
4593, NGC 526A, NGC 5506, NGC 5548, NGC 7172, NGC 7213, NGC 7314 and NGC 7469.
Fig.\ 1a shows their best-fit points in open symbols, and the corresponding
1-$\sigma$ confidence contours are shown in thin solid line. For clarity, the
inset shows the best-fit points (except that of MCG --6-30-15) without the
contours. The open circles denote RQ Seyferts from the sample of Nandra \&
Pounds (1994), and the open square and triangles, NGC 4151 and 4U 0241+61,
respectively.

We see an extremely strong correlation. At the hard end, objects with
$\Gamma\sim 1.5$ have almost no reflection. The reflection strength then
increases with increasing $\Gamma$. Reflection appears to saturate at $R\sim 2$
as our softest Seyfert, MCG --6-30-15, has $R\sim 2$ at the best-fit
$\Gamma\approx 2.4$, somewhat below an extrapolation of the trend for objects
with harder spectra. We also see that NGC 4151 and 4U 0241+61 show rather
average values of $\Gamma$ and $R$.

We checked in Section 3 below that the correlation is not an artefact of our
fitting procedure. Indeed, although there are intrinsic $R$-$\Gamma$
correlations due to finite measurement errors for each observation, we clearly
see in Fig.\ 1a that their extents are much less (especially for data with good
statistics) than the extent of the global correlation.  Objects with hard power
laws and significant reflection would have reflection easily measured by
\ginga\/ due to their high count rate at $\ga 10$ keV. Also, soft power laws
without reflection would also have been measured as such by \ginga, albeit with
relatively large errors. The absence of such objects in the sample strongly
supports the physical reality of the correlation.

We note that a similar $R(\Gamma)$ correlation has been observed for NGC
5548 (Magdziarz et al.\ 1998). We show the contours for this object alone in
Fig.\ 1b, including one contour omitted in Fig.\ 1a [due to our criterion of
$\Delta (R)\le 2$], which also obeys the correlation. We also find
an $R(\Gamma)$ correlation for Mkn 509.

We then consider hard-state X-ray binaries, for which fit results are shown by
heavy contours and filled symbols in Fig.\ 1a. The 12 filled squares correspond
to Cyg X-1 (at assumed $i=30\degr$). We see that Cyg X-1 has typically {\it
more\/} reflection than average for Seyferts with the same $\Gamma$ (although
its contours are still within the area covered by Seyferts). Although the
presence of an $R(\Gamma)$ correlation cannot be proven from these data alone,
Done \& \.Zycki (1999) found such a correlation in observations by \exosat\/
and \asca.

The filled circles in Figs.\ 1a, c correspond to the data for GX 339--4. We
fit the \ginga\/ data as in Zdziarski et al.\ (1998), assuming $i=45\degr$. We
see that the contours show a highly significant correlation themselves, as
found before by Ueda et al.\ (1994), and that their correlation is consistent
with that for average Seyferts.

Filled triangles in Fig.\ 1a correspond to the X-ray bursters (at assumed
$i=60\degr$). The large triangle and 2 small ones correspond to GS 1862--238
and 4U 1608--522, respectively, fitted with the same model as that used for
Seyferts.  We see that strong reflection is present at high statistical
significance in both objects, and that its strength is consistent with
that typical for Seyferts.

Finally, we consider 6 \ginga\/ observations of 5 nearby radio galaxies, 3C
111, 3C 382, 3C 390.3, 3C 445, Cen A, fitted as in Wo\'zniak et al.\ (1998).
Those authors find that they show significantly less reflection than the
average for RQ Seyfert 1s. Those data are shown by crosses in the inset of
Fig.\ 1a. We see that these data points still obey the overall correlation,
albeit with reflection indeed much weaker than that average for RQ Seyferts.

\section{STATISTICAL ANALYSIS}

The correlation between $R$ and $\Gamma$ was first tested using 2
rank-order correlation tests of Spearman and Kendall (Press et al.\ 1992). We
use our sample of RQ Seyferts but excluding MCG --6-30-15, for which
$R$ appears saturated. We find $r_{\rm S}=0.91$, $\tau_{\rm K}=0.76$
with the significance levels of $r_{\rm S}$ and $\tau_{\rm K}$ being $>0$ of
$< 10^{-18}$ and  $< 10^{-13}$, respectively. Thus, the correlation
between $R$ and $\Gamma$ is indeed very strong. Note, however, that those
statistical methods do not take into account individual measurement errors.

We have then fitted a few phenomenological functions to these data in order to
express quantitatively the dependence. The best model found is a power law,
$R  = u \Gamma^v$, giving $u = (1.4\pm 0.1)\times 10^{-4}$, $v=
12.4\pm 1.2$ at $\chi_\nu^2 = 45/44$.  Errors along both axes were taken into
account in the fitting as described in Brandt (1997).

As pointed out above, the error contours for individual measurements exhibit
themselves significant intrinsic correlations, which should be taken into
account. Direct estimates of the intrinsic correlation coefficients, $\rho$ (as
given by {\sc xspec}), are not reliable due to an asymmetry of the covariance
matrix for $(\Gamma,\, R)$. However, this asymmetry is reflected in
deviation of a contour from the shape of an ellipse with the axes parallel to
the coordinate axes, and we can thus estimate $\rho$ by graphically measuring
its shape. The obtained values vary from $\rho \sim 0.7$ for contours
with $R \sim 0$ to $\rho \sim 0.95$ for those with the strongest
reflection.

We then use the individual values of $\rho$ in fitting the power-law model
using a method of Brandt (1997). This yields $u= (1.7\pm 0.8) \times 10^{-4}$,
$v=11.9\pm 0.7$, which is within the 1-$\sigma$ confidence ranges of the $u$,
$v$ above. On the other hand, $\chi_\nu^2 = 115/44$ is substantially larger
than that of the fit above. This can be interpreted as due to the area of the
distorted ellipses being much smaller than that of rectangles obtained from
taking the errors on $\Gamma$ and $R$ as uncorrelated.

Note that the latter large value of $\chi_\nu^2$ reflects an actual spread of
values of $R$ for a given $\Gamma$ due to a range of physical conditions
in the objects rather than due to measurement errors. Still, $R$ is very
strongly correlated with $\Gamma$ even taking into account this physical width
of the correlation as well as the measurement-related correlation between the
parameters. This can be quantified by fitting a constant, $R = u$,
to the data, which corresponds to the null hypothesis of the absence of a
physical correlation. The fit is performed with the same method as above, i.e.,
taking into account the intrinsic $R(\Gamma)$ correlation. We obtain $u=
0.22\pm 0.02$ at $\chi_\nu^{2}= 312/45$. From comparison to $\chi_\nu^2$ for
the corresponding power-law dependence, the probability that there is no
correlation between $R$ and $\Gamma$ beyond the measurement-related one is
$2\times 10^{-10}$, as obtained from the F-test.

\section{THEORETICAL INTERPRETATION}

We note first that since at least some individual objects (NGC 5548, Mkn 509,
GX 339--4, Cyg X-1) exhibit an $R(\Gamma)$ correlation similar to the
global one, the correlation cannot be due to an orientation effect (e.g.\
hard objects with weak reflection being oriented edge-on). Rather, its cause
must be some feedback within the source in which the presence of a cold medium
(responsible for reflection) affects the hardness of the X-ray spectra. A
natural explanation for the feedback is that the cold medium emits soft photons
that irradiate the X-ray source and serve as seeds for Compton upscattering.
Then, the larger the effective solid angle subtended by the reflector, the
stronger the flux of soft photons, and, consequently, the stronger cooling of
the plasma. In the case of a thermal plasma, the larger the cooling by seed
photons incident on the plasma, the softer the resulting X-ray power-law
spectra. Indeed, X-ray and soft \g-ray spectra of both Seyferts and black-hole
binaries in the hard state appear to be due to Comptonization by thermal
electrons rather than non-thermal ones (Gierli\'nski et al.\ 1997; Johnson et
al.\ 1997; Grove et al.\ 1998; Zdziarski et al.\ 1996, 1997, 1998).

We consider here 2 models. In the first one, a central uniform, optically-thin,
sphere with unit radius and unit luminosity is surrounded by a flat,
optically-thick disc (e.g.\ Poutanen, Krolik \& Ryde 1997). The disc extends
down to a radius $d\geq 0$ (Fig.\ 2). In the limit of $d=0$, this geometry
corresponds to a localized active region above the surface of an
optically-thick accretion disc (Haardt, Maraschi \& Ghisellini 1994; Stern et
al.\ 1995). On the other hand, $d\ga 1$ corresponds to an inner hot accretion
disc (Shapiro, Lightman \& Eardley 1976), possibly advection-dominated (e.g.\
Abramowicz et al.\ 1995; Narayan \& Yi 1995; Zdziarski 1998) surrounded by an
optically-thick outer disc. We assume that the sphere is a hot plasma radiating
via thermal Comptonization and the seed photons for upscattering are the
reprocessed ones emitted by the cold disc.

In order to compute the fraction of the emission of the sphere intercepted by
the disc, we need the flux, $F(r)$, incident on the disc at a distance $r$
from the center. The specific intensity of radiation emitted by the sphere in a
direction towards the disc is $I=j l$, where the emission coefficient is
$j=3/(16\upi^2)$ for a uniform sphere with unit luminosity and unit radius, and
$l$ is the length through the sphere in this direction. Then $F(r) =j \int {\rm
d} \Omega\, l \cos \alpha \equiv 3h(r) /(16\upi^2)$, where $\alpha$ is the
angle between the direction of an incoming ray and the disc normal, and we
denoted the integral of $l$ as $h(r)$. By integration, we obtain,
\begin{eqnarray}\label{hr} \lefteqn{ h(r)=(4/3)\times} \nonumber\\ \lefteqn{
\cases{ \left[\left(2-r^{-2}\right)E\left(r^2\right)+
\left(r^{-2}-1\right)K\left(r^2\right)\right], & $r<1$;\cr
\left[\left(2r-r^{-1}\right)E\left(r^{-2}\right)+
2\left(r^{-1}-r\right)K\left(r^{-2}
\right)\right], & $r\geq 1$,\cr } } \end{eqnarray} where $E$ and $K$ are
complete elliptic integrals. The function $h(r)$ has the following properties:
$h(0)=\upi$, $h(1)=4/3$, $h(r\!\gg\! 1) \rightarrow \upi/ (4 r^3)$,
$\int_1^\infty {\rm d}r\, r h(r)=8/9$, $\int_0^\infty {\rm d}r\, r
h(r)=2\upi/3$, and $\int_0^\infty {\rm d}r\, r [h(r)]^2=\upi^2/3$. The
luminosity of the disc is obtained by integrating the flux over the surface,
$L_{\rm d}(d)=4\upi \int_d^\infty {\rm d}r\, r F(r)$. In particular, $L_{\rm
d}(0)=1/2$ (as expected from the isotropy of the sphere emission) and $L_{\rm
d}(1)= 2/(3\upi)$ (as found by Chen \& Halpern 1989). The ratio of the
luminosity of the disc to the part of the sphere luminosity radiated outward
can be identified with the average strength of Compton reflection,
$R = L_{\rm d}/(1-L_{\rm d})$.

\begin{figure}
\begin{center} \leavevmode \epsfxsize=7cm \epsfbox{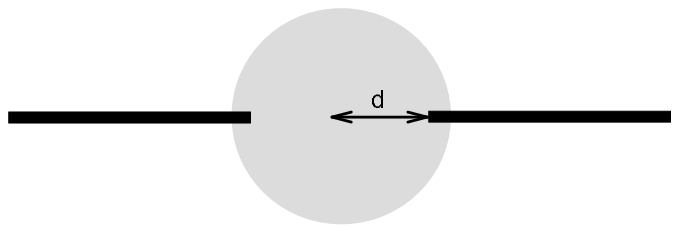} \end{center}
\caption{The geometry with a central hot plasma surrounded by a cold disc, see
Section 4. } \end{figure}

The reflection albedo is relatively low, $a\sim 0.1$--0.2 (e.g., Magdziarz \&
Zdziarski 1995), and most of the incident flux is reradiated isotropically as
soft blackbody radiation with the specific intensity $I(r)=\eta F(r)/\upi$,
where $\eta\equiv 1-a$.  If the disc radiates solely the reprocessed flux,
$I(r)= 3\eta h(r)/(16\upi^3)$.  The power in seed photons scattered in the
sphere (assuming a unit optical depth) is then, \begin{equation}\label{ldisk}
L_{\rm s}(d) \approx 4\upi \int_d^\infty {\rm d}r\, r I(r) h(r)= {3\eta \over
4\upi^2} \int_d^\infty {\rm d}r\, r h^2(r), \end{equation} where $L_{\rm
s}(0)\approx\eta/4$, and $A(d)\equiv 1/L_{\rm s}(d)$ is the amplification factor
of the process of thermal Comptonization.

We then use an estimate of Beloborodov (1999a) for $\Gamma(A)$, $\Gamma\approx
2.33(A-1)^{-\delta}$, where $\delta \approx 1/6$, 1/10 for X-ray binaries and
AGNs, respectively. From $\Gamma[A(d)]$ and $R(d)$ above, we obtain the model
prediction for $R(\Gamma)$, plotted for $\delta=1/10$ and $a=0.15$ in a dashed
curve in the inset of Fig.\ 1a. We see that this model can reproduce the data
in a middle range of $\Gamma$ and $R$ (especially for black-hole binaries), but
it cannot reproduce $R>1$. In fact, scattering of reflected photons in the hot
plasma (neglected in the above model) will significantly reduce reflection for
small $d$, i.e., for large $\Gamma$, which reduces the maximum possible $R$ to
a value $<1$ (e.g.\ Poutanen et al.\ 1997). (Also, intrinsic dissipation in the
disc will reduce $R$ for given $\Gamma$.) Disc flaring (Shakura \& Sunyaev
1973; enhanced by irradiation, Vrtilek et al.\ 1990) and the presence of
Thomson-thick molecular torii in some Seyferts can increase $R$. However, such
torii should also be present in some Seyferts with $\Gamma\la 1.6$, contrary to
our data showing $R\ll 1$. Concluding, this model may apply to some sources but
it is unlikely to account for the full range of $R$ observed in Seyferts.

Another model to explain the observed range of $\Gamma$ and $R$ has
been proposed by Beloborodov (1999a,  b). In this model, the parameter
controlling the spectral shape is the bulk-motion velocity of the X-ray
emitting plasma located above an accretion disc. A mildly relativistic outflow
reduces the downward flux, which in turn reduces both reflection and
reprocessing in the disc. The reduction of the reprocessed flux incident on the
hot plasma leads to a spectral hardening (analogously to the previous model).
Such outflows were also proposed by Wo\'zniak et al.\ (1998) to explain the
weakness of reflection in radio galaxies. On the other hand, a mildly
relativistic motion directed towards the disc will enhance reflection and
cooling, leading to soft spectra.

Beloborodov [1999a, eqs.\ (3) and (7)] derives approximate formulae for $R$ and
$A$ as functions of the bulk motion velocity. We show here the same model as in
Beloborodov (1999b), for $\delta=1/10$, $i=45\degr$ and his geometry-dependent
factor, $\mu_{\rm s}=0.55$, by solid curve in the inset of Fig.\ 1a. This model
reproduces well the observed average $R(\Gamma)$. Variations in the energy of
the seed photons, the plasma temperature and optical depth (affecting
$\delta$), plasma geometry and orientation from source to source can reproduce
the width of the correlation. Thus, this model appears fully capable to explain
the observed correlation.

\section{DISCUSSION}

Our main finding is of a very strong correlation between the intrinsic spectral
slope and the relative strength of Compton reflection in Seyferts. The
statistical significance of the correlation is $1-2\times 10^{-10}$ after
removing the effect of an intrinsic correlation between those two quantities in
each individual measurement. We have also found that the correlation is
satisfied by hard-state spectra of 4 X-ray binaries containing either black
holes or weakly-magnetized neutron stars. Since that sample is limited, more
data are needed to test the generality of the correlation in X-ray binaries.

Our general interpretation of the correlation is as follows. The X-ray and soft
\g-ray spectra of these classes of sources are well modeled by thermal
Comptonization in hot plasmas, which emit power-law X-ray spectra with the
slope related to the rate of cooling by incident soft photons. Then, the
spectral slope will be correlated with the strength of reflection provided the
main source of the cooling photons is emission of the same medium that is
responsible for the observed reflection. A specific model with a mildly
relativistic bulk motion of the hot plasma above an accretion disk (Beloborodov
1999a) can quantitatively explain the correlation.

An observational prediction of our interpretation of the correlation in terms
of plasma cooling is that objects with soft X-ray spectra should on average
have lower electron temperatures than those with hard X-ray spectra. This
effect might, however, be compensated for by a reduction of the temperature due
to \ee\ pair production, which is more efficient in hotter plasmas.

The presence of the correlation strongly argues against thermal synchrotron
radiation being a significant source of seed photons for Comptonization in
those sources. This agrees with modeling of that process showing it to be
insignificant in luminous sources (Zdziarski et al.\ 1998; Wardzi\'nski \&
Zdziarski, in preparation).

An interesting issue is whether black-hole binaries in the soft state obey the
correlation. Their spectra in hard X-rays are typically power laws with
$\Gamma\sim 2$--3 (Grove et al.\ 1998). No systematic studies of the presence
of reflection in this state exist yet. In the case of Cyg X-1, Gierli\'nski et
al.\ (1999) find $R < 1$, which is less than that predicted by the correlation
found here. This can be explained by hard X-ray spectra in the soft state being
due to emission by non-thermal plasmas (Poutanen \& Coppi 1998; Gierli\'nski et
al.\ 1999), in which case an $R(\Gamma)$ correlation is {\it not\/} expected
theoretically. This is because Compton scattering by non-thermal electrons
leads to a photon spectrum with the slope related to the slope of the electron
distribution rather than to the cooling rate.

\section*{ACKNOWLEDGMENTS}

We thank J. Poutanen for valuable comments and suggestions, K. Ebisawa for
providing us with his data on Cyg X-1, and A. Beloborodov, M. Gierli\'nski and
G. Wardzi\'nski for discussions. This research has been supported in part by
the KBN grants 2P03C00511p0(1,4) and 2P03D00614 and NASA grants and contracts,
and it has made use of data obtained from the Leicester Database and Archive
Service at the Dept.\ of Physics and Astronomy, Leicester Univ., UK.

\bsp

\label{lastpage}

\end{document}